\documentclass[a4,11pt]{amsart}
\usepackage{amssymb}
\usepackage{amsmath}
\usepackage{amsfonts}
\usepackage{amsmath,amssymb,pstricks,pst-plot,pst-node,psfrag,graphicx,pst-grad,amsfonts}
\usepackage{float}

\begin{document}

\title{Transmuted Lindley-Geometric Distribution and its applications}

\author[Faton Merovci]{Faton Merovci  }
\address{Faton Merovci
\newline \indent Department of Mathematics,
\newline \indent University of Prishtina "Hasan Prishtina",
\newline \indent Republic of Kosovo}

\email{fmerovci@yahoo.com}

\author[Ibrahim Elbatal]{Ibrahim Elbatal  }
\address{Ibrahim Elbatal
\newline \indent Institute of Statistical Studies and Research,
\newline \indent Department of Mathematical Statistics,
\newline \indent Cairo University}

\email{i\_elbatal@staff.cu.edu.eg}

\maketitle

\begin{abstract}

A functional composition of the cumulative distribution function of one
probability distribution with the inverse cumulative distribution function
of another is called the transmutation map. In this article, we will use the
quadratic rank transmutation map (QRTM) in order to generate a flexible
family of probability distributions taking Lindley geometric distribution as
the base value distribution by introducing a new parameter that would offer
more distributional flexibility. It will be shown that the analytical
results are applicable to model real world data.

\end{abstract}

\section{Introduction and Motivation}

The Lindley distribution was originally proposed by Lindley \cite{lindley} in the context of Bayesian statistics, as a counter example of fudicial statistics. More details on the Lindley distribution can be found in Ghitany et al. \cite{gitany}.\\

A random variable X is said to have the Lindley distribution with  parameter $\theta$   if its probability density is defined as
\begin{equation}\label{eq1.1}
f_{L}(x,\theta )=\frac{\theta ^{2}}{\theta +1}(1+x)e^{-\theta x}\text{ };%
\text{ }x>0,\theta >0,
\end{equation}%

The corresponding cumulative distribution function (c.d.f.) is:
\begin{equation}\label{eq1.2}
F_{L}(x,\theta )=1-(1+\frac{\theta x}{\theta +1})e^{-\theta x},x>0,\theta >0.
\end{equation}%

Many authors gives generalized Linldey distribution like Sankaran \cite{sankaran} introduced the discrete Poisson- Lindley, Mahmoudi and Zakerzadeh \cite{mahmuti} introduced generalized Lindley distribution,Bakouch et al. \cite{bakouch} introduced extended Lindley (EL)
distribution, Adamidis and Loukas \cite{adamis} introduced 
exponential geometric (EG) distribution.

Recently, Hojjatollah and Mahmoudi \cite{2012} introduced  Lindley- Geometric distribution where the cdf and pdf
of this distribution are given by
\begin{equation}\label{eq1.3}
F_{LG}(x,\theta ,p)=\frac{1-(1+\frac{\theta x}{\theta +1})e^{-\theta x}}{
1-p(1+\frac{\theta x}{\theta +1})e^{-\theta x}},x>0,\theta >0,0<p<1,
\end{equation}
and
\begin{equation}\label{eq1.4}
f_{LG}(x,\theta ,p)=\frac{\theta ^{2}}{\theta +1}(1-p)(1+x)e^{-\theta x}%
\left[ 1-p(1+\frac{\theta x}{\theta +1})e^{-\theta x}\right] ^{-2},
\end{equation}
respectively. In this paper, we introduce a new lifetime distribution by
transmuted and compounding Lindley and geometric distributions named
transmuted Lindley geometric distribution. The concept of transmuted
explained in the following subsection.

\subsection{Transmutation Map}

In this subsection we demonstrate transmuted probability distribution. Let $F_{1}$ and $F_{2}$ be the cumulative distribution functions, of two
distributions with a common sample space. The general rank transmutation  is defined as
\begin{equation*}
G_{R12}(u)=F_{2}(F_{1}^{-1}(u))\text{ \ and }G_{R21}(u)=F_{1}(F_{2}^{-1}(u)).
\end{equation*}
Note that the inverse cumulative distribution function also known as
quantile function is defined as
\begin{equation*}
F^{-1}(y)=\text{ inf}_{x\in R}\left\{ F(x)\geq y\right\} \text{ for }y\in \left[ 0,1\right] .
\end{equation*}%
The functions $G_{R12}(u)$ and $G_{R21}(u)$ both map the unit interval $I=$ $%
\left[ 0,1\right] $ into itself, and under suitable assumptions are mutual
inverses and they satisfy $G_{Rij}(0)=0$ and $G_{Rij}(0)=1.$ A quadratic Rank
Transmutation Map (QRTM) is defined as%
\begin{equation}
G_{R12}(u)=u+\lambda u(1-u),\left\vert \lambda \right\vert \leq 1,  \tag{1.3}
\end{equation}%
from which it follows that the cdf's satisfy the relationship%
\begin{equation}\label{eq1.4}
F_{2}(x)=(1+\lambda )F_{1}(x)-\lambda F_{1}(x)^{2}
\end{equation}%
which on differentiation yields,%
\begin{equation}\label{eq1.5}
f_{2}(x)=f_{1}(x)\left[ (1+\lambda )-2\lambda F_{1}(x)\right]
\end{equation}%
where $f_{1}(x)$ and $f_{2}(x)$ are the corresponding pdfs associated with
cdf $F_{1}(x)$ and $F_{2}(x)$ respectively. An extensive information about
the quadratic rank transmutation map is given in Shaw et al. (2007). Observe
that at $\lambda =0$ we have the distribution of the base random variable.
The following lemma proved that the function $f_{2}(x)$ in given \eqref{eq1.5}
satisfies the property of probability density function.\medskip \newline
\textbf{Lemma: }$f_{2}(x)$ given in \eqref{eq1.5} is a well defined probability
density function.\medskip \newline
\textbf{Proof. }\medskip \newline
Rewriting $f_{2}(x)$ as $f_{2}(x)=f_{1}(x)\left[ (1-\lambda (2F_{1}(x)-1)\right] $
 we observe that $f_{2}(x)$ is nonnegative. We need to show that
the integration over the support of the random variable is equal one.
Consider the case when the support of $f_{1}(x)$ is $(-\infty ,\infty )$. In
this case we have%
\begin{align*}
\int\limits_{-\infty }^{\infty }f_{2}(x)dx &=\int\limits_{-\infty
}^{\infty }f_{1}(x)\left[ (1+\lambda )-2\lambda F_{1}(x)\right] dx \\
&=(1+\lambda )\int\limits_{-\infty }^{\infty }f_{1}(x)dx-\lambda
\int\limits_{-\infty }^{\infty }2f_{1}(x)F_{1}(x)dx \\
&=(1+\lambda )-\lambda \\
&=1
\end{align*}%
Similarly, other cases where the support of the random variable is a part of
real line follows. Hence $f_{2}(x)$ is a well defined probability density
function. We call $f_{2}(x)$ the transmuted probability density of a random
variable with base density $f_{1}(x)$. Also note that when $\lambda =0$ then
$f_{2}(x)=f_{1}(x).$ This proves the required result.

Many authors dealing with the generalization of some well- known
distributions. Aryal and Tsokos (2009) defined the transmuted generalized
extreme value distribution and they studied some basic mathematical
characteristics of transmuted Gumbel probability distribution and it has
been observed that the transmuted Gumbel can be used to model climate data.
Also Aryal and Tsokos (2011) presented a new generalization of Weibull
distribution called the transmuted Weibull distribution . Recently, Aryal
(2013) proposed and studied the various structural properties of the
transmuted Log- Logistic distribution, and Muhammad khan and king (2013)
introduced the transmuted modified Weibull distribution which extends recent
development on transmuted Weibull distribution by Aryal et al. (2011), Merovci \cite{faton1},\cite{faton2},\cite{faton3}introduced the transmuted Rayleigh distribution, transmuted generalized Rayleigh distribution, transmuted Lindley distribution and
they studied the mathematical properties and maximum likelihood estimation
of the unknown parameters.

\subsection{Transmuted Lindley Geometric Distribution}

In this section we studied the transmuted Lindley geometric (TLG)
distribution . Now using \eqref{eq1.3}and \eqref{eq1.4} we have the cdf of transmuted
Lindley geometric (TLG) distribution%
\begin{equation}\label{eq2.1}
F_{TLG}(x,\theta ,p,\lambda )=\frac{1-(1+\frac{\theta x}{\theta +1}
)e^{-\theta x}}{1-p(1+\frac{\theta x}{\theta +1})e^{-\theta x}}\left[
1+\lambda -\lambda \left( \frac{1-(1+\frac{\theta x}{\theta +1})e^{-\theta x}
}{1-p(1+\frac{\theta x}{\theta +1})e^{-\theta x}}\right) \right]
\end{equation}
where $\lambda $ is the transmuted parameter. The corresponding probability
density function (pdf) of the transmuted Lindley geometric is given by%
\begin{align}\label{eq2.2}
f_{TLG}(x,\theta ,p,\lambda ) &=f_{LG}(x)\left[ (1+\lambda )-2\lambda
F_{LG}(x)\right]  \notag \\
&=\frac{\theta ^{2}}{\theta +1}(1-p)(1+x)e^{-\theta x}\left[ 1-p(1+\frac{
\theta x}{\theta +1})e^{-\theta x}\right] ^{-2}  \notag \\
&\times \left\{ (1+\lambda )-2\lambda \left( \frac{1-(1+\frac{\theta x}{
\theta +1})e^{-\theta x}}{1-p(1+\frac{\theta x}{\theta +1})e^{-\theta x}}
\right) \right\} ,
\end{align}
respectively.

Figure 1 and figure 2 illustrates some of the possible shapes of the pdf and cdf of TLG distribution  for selected values of the parameters $\theta, p $ and $\lambda$, respectively.
me
\begin{figure}[H]
\centering
  \includegraphics[width=8.0cm]{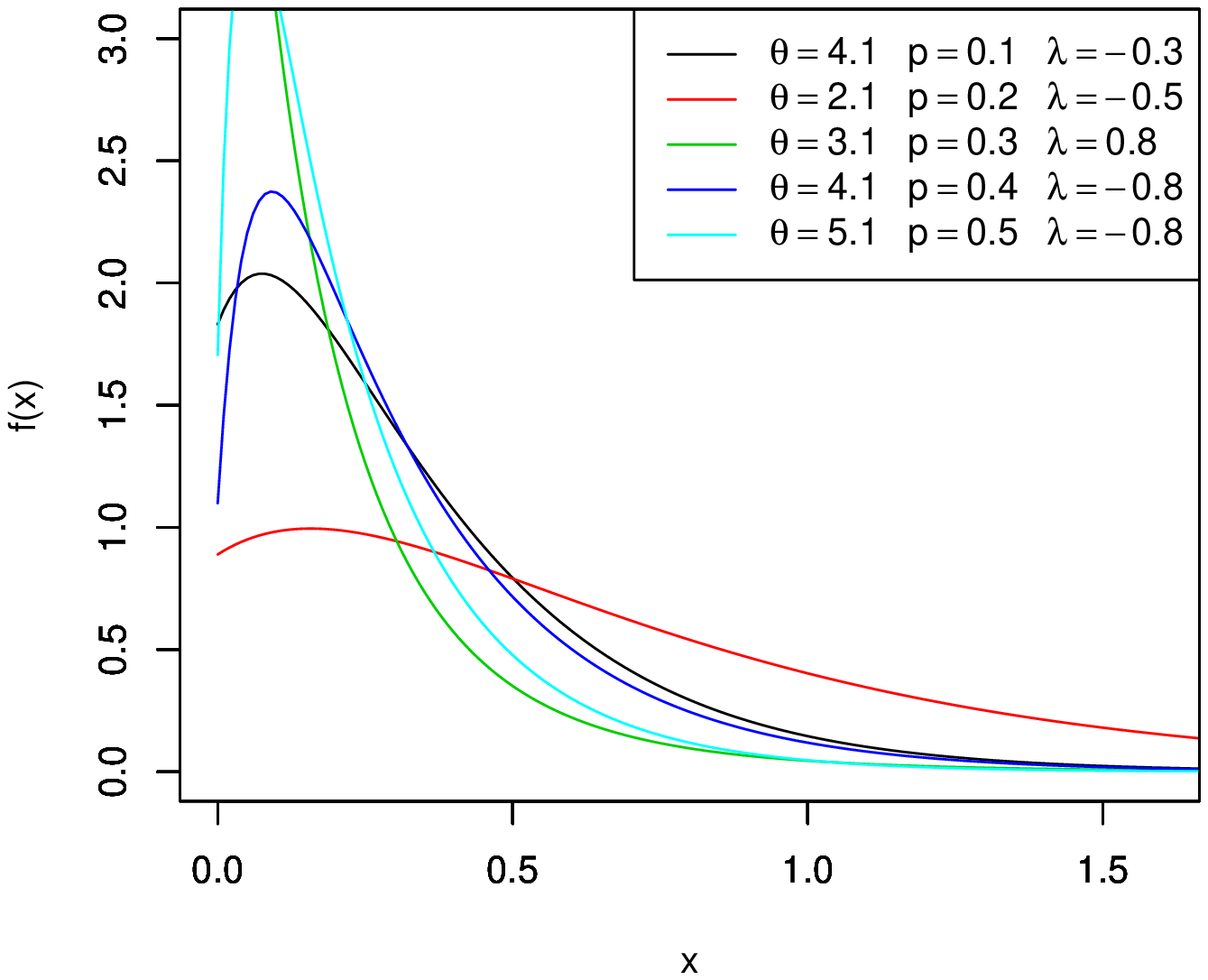}
 \caption{The pdf's of various TLG distributions.\label{fig1.pdf}}
\end{figure}
 \begin{figure}[H]
\centering
  \includegraphics[width=8.0cm]{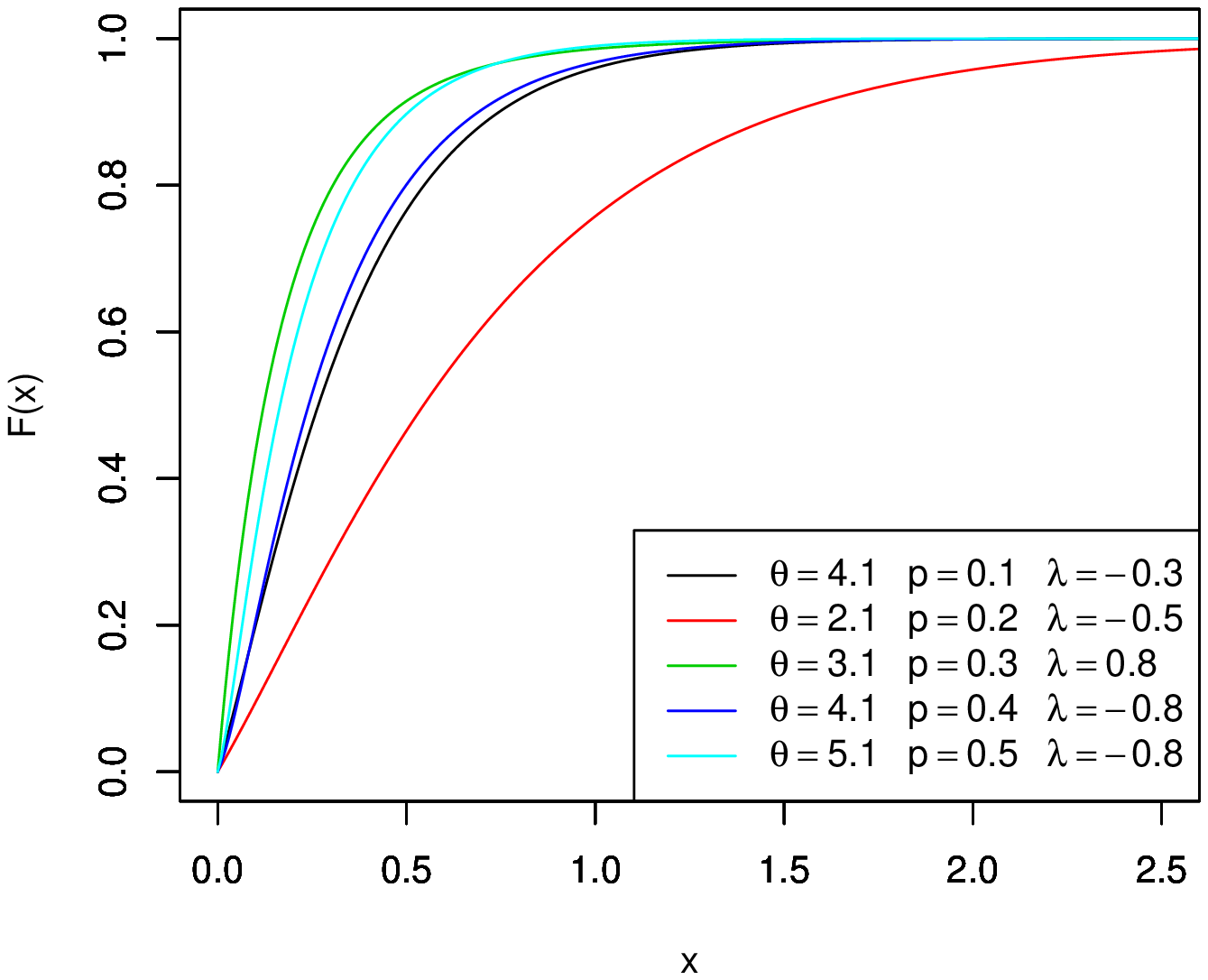}
 \caption{The cdf's of various TLG distributions.\label{fig2.pdf}}
 \end{figure}

 The reliability function $(RF)$ of the transmuted Lindley
geometric distribution is denoted by $R_{TLG}(x)$ also known as the survivor
function and is defined as%
\begin{align}\label{eq2.3}
R_{TLG}(x) &=1-F_{TLG}(x)  \notag \\
&=1-\frac{1-(1+\frac{\theta x}{\theta +1})e^{-\theta x}}{1-p(1+\frac{\theta
x}{\theta +1})e^{-\theta x}}\left[ 1+\lambda -\lambda \left( \frac{1-(1+
\frac{\theta x}{\theta +1})e^{-\theta x}}{1-p(1+\frac{\theta x}{\theta +1}
)e^{-\theta x}}\right) \right] .
\end{align}
Figure 3 illustrates some of the possible shapes of the survival function of transmuted Lindley geometric distribution  for selected values of the parameters $\theta, p$ and $\lambda$, respectively.
\begin{figure}[H]
\centering
  \includegraphics[width=8.0cm]{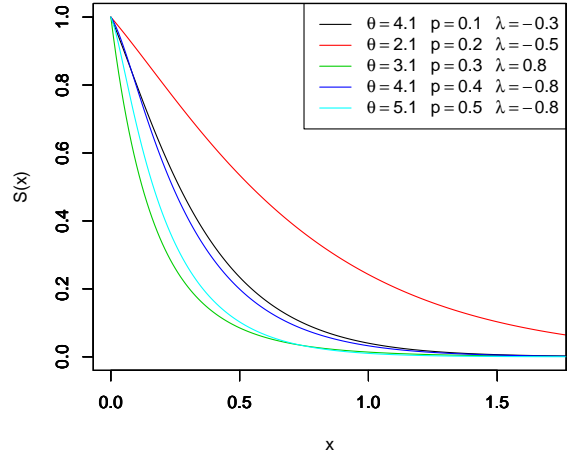}
 \caption{The survival function  of various transmuted Lindely geometric distributions.\label{fig3.pdf}}
 \end{figure}
It is important to note that $R_{TLG}(x)+F_{TLG}(x)=1$ . One of the
characteristic in reliability analysis is the hazard rate function (HF)
defined by%
\begin{equation}\label{eq2.4}
h_{TLG}(x)=\frac{f_{TLG}(x)}{1-F_{TLG}(x)}
\end{equation}%

Figure 4 illustrates some of the possible shapes of the hazard function of transmuted Lindley geometric distribution  for selected values of the parameters $\theta, p$ and $\lambda$, respectively.
\begin{figure}[H]
\centering
  \includegraphics[width=8.0cm]{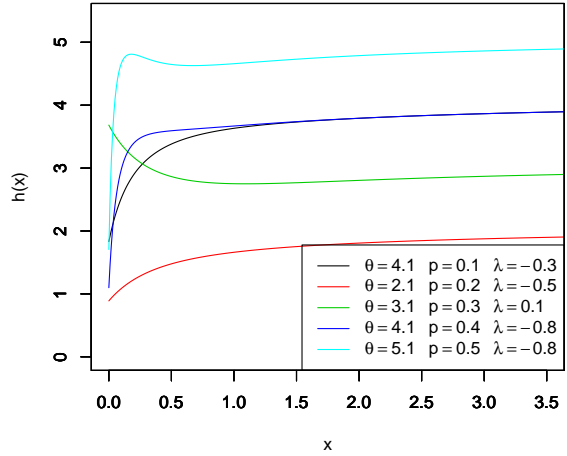}
 \caption{The survival function  of various transmuted Lindely geometric distributions.\label{fig3.pdf}}
 \end{figure}
It is important to note that the units for $h_{TLG}(x)$ is the probability
of failure per unit of time, distance or cycles. These failure rates are
defined with different choices of parameters.The cumulative hazard function
of the transmuted Lindley geometric distribution is denoted by $H_{TLG}(x)$
and is defined as%
\begin{equation}\label{eq2.5}
H_{TLG}(x)=-\ln \left\vert \frac{1-(1+\frac{\theta x}{\theta +1})e^{-\theta
x}}{1-p(1+\frac{\theta x}{\theta +1})e^{-\theta x}}\left[ 1+\lambda -\lambda
\left( \frac{1-(1+\frac{\theta x}{\theta +1})e^{-\theta x}}{1-p(1+\frac{
\theta x}{\theta +1})e^{-\theta x}}\right) \right] \right\vert
\end{equation}
It is important to note that the units for $H_{TLG}(x)$ is the cumulative
probability of failure per unit of time, distance or cycles. we can show
that . For all choice of parameters the distribution has the decreasing
patterns of cumulative instantaneous failure rates.

\section{Statistical Properties}

This section is devoted to studying statistical properties of the $(TLG)$
distribution.

\subsection{Moments}

In this subsection we discuss the $r_{th}$ moment for $(TLG)$ distribution.
Moments are necessary and important in any statistical analysis, especially
in applications. It can be used to study the most important features and
characteristics of a distribution (e.g., tendency, dispersion, skewness and
kurtosis).\medskip \newline
\textbf{Theorem (3.1)}.\medskip \newline
If $X$ has $TLG$ $(\Phi ,x)$ $,\Phi =(\theta ,p,\lambda )$ then the $r_{th}$
moment of $X$ is given by the following
\begin{align}\label{eq3.1}
\mu _{r}^{^{\prime }}(x) &=A_{ig}\frac{\Gamma (r+i+1)}{\left( \theta
(j+1)\right) ^{r+i+1}}\left[ 1+\frac{r+i+1}{\left( \theta (j+1)\right) }%
\right]  \notag \\
&-B_{ij}\left\{ \frac{\Gamma (r+i+1)}{\left( \theta (j+1)\right) ^{r+i+1}}%
\left[ 1+\frac{r+i+1}{\left( \theta (j+1)\right) }\right] -\frac{\Gamma
(r+i+1)}{\left( \theta (j+2)\right) ^{r+i+1}}\left[ 1+\frac{r+i+1}{\left(
\theta (j+2)\right) }\right] \right.  \notag \\
&-\left. \frac{\theta }{\theta +1}\left( \frac{\Gamma (r+i+2)}{\left(
\theta (j+2)\right) ^{r+i+2}}\right) \left[ 1+\frac{r+i+2}{\left( \theta
(j+2)\right) }\right] \right\},
\end{align}
where
\begin{equation*}
A_{ig}=\frac{\theta ^{2}(1+\lambda )}{\theta +1}(1-p)\sum\limits_{j=0}^{%
\infty }\sum\limits_{i=0}^{j}\binom{j}{i}(j+1)p^{j}\left( \frac{\theta }{%
\theta +1}\right) ^{i},
\end{equation*}
and
\begin{equation*}
B_{ij}=\frac{\lambda \theta ^{2}}{\theta +1}(1-p)\sum\limits_{j=0}^{\infty
}\sum\limits_{i=0}^{j}\binom{j}{i}(j+1)(j+2)p^{j}\left( \frac{\theta }{
\theta +1}\right) ^{i}.
\end{equation*}
\textbf{Proof}:

Let $X$ be a random variable with density function \eqref{eq2.2}. The $r_{th}$
ordinary moment of the $(TLG)$ distribution is given by
\begin{align}\label{eq3.2}
\mu _{r}^{^{\prime }}(x) &=E(X^{r)}=\int\limits_{0}^{\infty
}x^{r}f(x,\Phi )dx  \notag \\
&=\frac{\theta ^{2}(1+\lambda )}{\theta +1}(1-p)\int\limits_{0}^{\infty
}(x^{r}+x^{r+1})e^{-\theta x}\left[ 1-p(1+\frac{\theta x}{\theta +1}
)e^{-\theta x}\right] ^{-2}dx  \notag \\
&-\frac{2\lambda \theta ^{2}}{\theta +1}(1-p)\int\limits_{0}^{\infty
}(x^{r}+x^{r+1})e^{-\theta x}\left( 1-(1+\frac{\theta x}{\theta +1}
)e^{-\theta x}\right) \left[ 1-p(1+\frac{\theta x}{\theta +1})e^{-\theta x}
\right] ^{-3}dx.
\end{align}
using the series expansion
\begin{equation}\label{eq3.3}
(1-z)^{-k}=\sum\limits_{j=0}^{\infty }\frac{\Gamma (k+j)}{\Gamma (k)j!}
z^{j},
\end{equation}
where$\left\vert z\right\vert <1$ and $k>0.$

 Equation \eqref{eq3.2} can be
demonstrated by
\begin{align}\label{eq3.4}
\mu _{r}^{^{\prime }}(x) &=\frac{\theta ^{2}(1+\lambda )}{\theta +1}
(1-p)\sum\limits_{j=0}^{\infty }(j+1)p^{j}\int\limits_{0}^{\infty
}(x^{r}+x^{r+1})(1+\frac{\theta x}{\theta +1})^{j}e^{-\theta (j+1)x}dx
\notag \\
&-\left\{ \frac{\lambda \theta ^{2}}{\theta +1}(1-p)\sum\limits_{j=0}^{
\infty }(j+1)(j+2)p^{j}\right.  \notag \\
&\left. \int\limits_{0}^{\infty }(x^{r}+x^{r+1})(1+\frac{\theta x}{
\theta +1})^{j}\left( 1-(1+\frac{\theta x}{\theta +1})e^{-\theta x}\right)
e^{-\theta (j+1)x}dx\right\} ,
\end{align}
also applying the binomial expression for $(1+\frac{\theta x}{\theta +1}%
)^{j} $ where
\begin{equation}\label{eq3.5}
(1+\frac{\theta x}{\theta +1})^{j}=\sum\limits_{i=0}^{j}\binom{j}{i}\left(
\frac{\theta }{\theta +1}\right) ^{i}x^{i},
\end{equation}%
substituting from \eqref{eq3.5} into \eqref{eq3.4} we get
\begin{align*}
\mu _{r}^{^{\prime }}(x) &=\left\{ \frac{\theta ^{2}(1+\lambda )}{\theta +1}
(1-p)\sum\limits_{j=0}^{\infty }\sum\limits_{i=0}^{j}\binom{j}{i}
(j+1)p^{j}\left( \frac{\theta }{\theta +1}\right) ^{i}\right. \\
&\left. \int\limits_{0}^{\infty }(x^{r+i}+x^{r+i+1})e^{-\theta
(j+1)x}dx\right\} \\
&-\left\{ \frac{\lambda \theta ^{2}}{\theta +1}(1-p)\sum\limits_{j=0}^{
\infty }\sum\limits_{i=0}^{j}\binom{j}{i}(j+1)(j+2)p^{j}\left( \frac{
\theta }{\theta +1}\right) ^{i}\right. \\
&\left. \int\limits_{0}^{\infty }(x^{r+i}+x^{r+i+1})\left( 1-(1+\frac{
\theta x}{\theta +1})e^{-\theta x}\right) e^{-\theta (j+1)x}dx\right\} \\
&=A_{ig}I_{1}-B_{ij}I_{2}
\end{align*}
where
\begin{equation*}
A_{ig}=\frac{\theta ^{2}(1+\lambda )}{\theta +1}(1-p)\sum\limits_{j=0}^{
\infty }\sum\limits_{i=0}^{j}\binom{j}{i}(j+1)p^{j}\left( \frac{\theta }{
\theta +1}\right) ^{i},
\end{equation*}
\begin{equation*}
B_{ij}=\frac{\lambda \theta ^{2}}{\theta +1}(1-p)\sum\limits_{j=0}^{\infty
}\sum\limits_{i=0}^{j}\binom{j}{i}(j+1)(j+2)p^{j}\left( \frac{\theta }{
\theta +1}\right) ^{i},
\end{equation*}
\begin{align*}
I_{1} &=\int\limits_{0}^{\infty }(x^{r+i}+x^{r+i+1})e^{-\theta (j+1)x}dx
\\
&=\frac{\Gamma (r+i+1)}{\left( \theta (j+1)\right) ^{r+i+1}}+\frac{\Gamma
(r+i+2)}{\left( \theta (j+1)\right) ^{r+i+2}} \\
&=\frac{\Gamma (r+i+1)}{\left( \theta (j+1)\right) ^{r+i+1}}\left[ 1+\frac{%
r+i+1}{\left( \theta (j+1)\right) }\right] ,
\end{align*}
and
\begin{align*}
I_{2} &=\int\limits_{0}^{\infty }(x^{r+i}+x^{r+i+1})\left( 1-(1+\frac{%
\theta x}{\theta +1})e^{-\theta x}\right) e^{-\theta (j+1)x}dx \\
&=\frac{\Gamma (r+i+1)}{\left( \theta (j+1)\right) ^{r+i+1}}\left[ 1+\frac{%
r+i+1}{\left( \theta (j+1)\right) }\right] -\frac{\Gamma (r+i+1)}{\left(
\theta (j+2)\right) ^{r+i+1}}\left[ 1+\frac{r+i+1}{\left( \theta
(j+2)\right) }\right] \\
&-\frac{\theta }{\theta +1}\left( \frac{\Gamma (r+i+2)}{\left( \theta
(j+2)\right) ^{r+i+2}}\right) \left[ 1+\frac{r+i+2}{\left( \theta
(j+2)\right) }\right] ,
\end{align*}
thus the $r_{th}$ moment is given by%
\begin{equation*}
\mu _{r}(x)=\theta \alpha ^{2}\sum\limits_{j=0}^{\infty
}\sum\limits_{m=0}^{\infty }(-1)^{j}\binom{j}{m}\alpha ^{m}\frac{\Gamma
(r+m+2)}{\left( \alpha (j+1)\right) ^{r+m+2}}\left[ (1+\lambda )\binom{%
\theta -1}{j}-2\lambda \tbinom{2\theta -1}{j}\right] .
\end{equation*}
Which completes the proof .\medskip \newline
We notice that if we put $\lambda =0,$ we get the $r_{th}$ moment of Lindley
geometric ( see Hojjatollah and Mahmoudi (2012)). Based on the first four
moments of the $(TLG)$ distribution, the measures of skewness $A(\Phi )$ and
kurtosis $k(\Phi )$ of the $(TLG)$ distribution can obtained as%
\begin{equation*}
A(\Phi )=\frac{\mu _{3}(\theta )-3\mu _{1}(\theta )\mu _{2}(\theta )+2\mu
_{1}^{3}(\theta )}{\left[ \mu _{2}(\theta )-\mu _{1}^{2}(\theta )\right] ^{%
\frac{3}{2}}},
\end{equation*}%
and
\begin{equation*}
k(\Phi )=\frac{\mu _{4}(\theta )-4\mu _{1}(\theta )\mu _{3}(\theta )+6\mu
_{1}^{2}(\theta )\mu _{2}(\theta )-3\mu _{1}^{4}(\theta )}{\left[ \mu
_{2}(\theta )-\mu _{1}^{2}(\theta )\right] ^{2}}.
\end{equation*}

\subsection{Moment Generating function}

In this subsection we derived the moment generating function of $(TLG)$
distribution.\medskip \newline

\textbf{Theorem (3.2)}: If $X$ has $(TLG)$ distribution, then the moment
generating function $M_{X}(t)$ has the following form%
\begin{align}\label{eq3.6}
M_{X}(t)&=\frac{A_{ig}\Gamma (i+1)}{\left( \theta (j+1)-t\right) ^{i+1}}\left[ 1+%
\frac{i+1}{\left( \theta (j+1)-t\right) }\right]  \notag \\
&-B_{ij}\left\{ \frac{\Gamma (i+1)}{\left( \theta (j+1)-t\right) ^{i+1}}%
\left[ 1+\frac{i+1}{\left( \theta (j+1)-t\right) }\right] \right.  \notag \\
&\left. -\frac{\Gamma (i+1)}{\left( \theta (j+2)-t\right) ^{i+1}}\left[ 1+%
\frac{i+1}{\left( \theta (j+2)-t\right) }\right] \right.  \notag \\
&\left. -\frac{\theta }{\theta +1}\left( \frac{\Gamma (i+2)}{\left( \theta
(j+2)-t\right) ^{i+2}}\right) \left[ 1+\frac{i+2}{\left( \theta
(j+2)-t\right) }\right] \right\}
\end{align}%
\textbf{Proof}.\medskip \newline

We start with the well known definition of the moment generating function
given by%
\begin{align}\label{eq3.7}
M_{X}(t) &=E(e^{tx})=\int\nolimits_{0}^{\infty }e^{tx}f_{TLG}(x,\Phi )dx
\notag \\
&=\frac{\theta ^{2}(1+\lambda )}{\theta +1}(1-p)\int\limits_{0}^{\infty
}(1+x)e^{-x(\theta -t)}\left[ 1-p(1+\frac{\theta x}{\theta +1})e^{-\theta x}
\right] ^{-2}dx  \notag \\
&-\frac{2\lambda \theta ^{2}}{\theta +1}(1-p)\int\limits_{0}^{\infty
}(1+x)e^{-x(\theta -t)}\left( 1-(1+\frac{\theta x}{\theta +1})e^{-\theta
x}\right) \left[ 1-p(1+\frac{\theta x}{\theta +1})e^{-\theta x}\right]
^{-3}dx.
\end{align}%
substituting from \eqref{eq3.3} and \eqref{eq3.5} into \eqref{eq3.7} we get
\begin{align}\label{eq3.8}
M_{X}(t) &=A_{ig}\int\limits_{0}^{\infty }(x^{i}+x^{i+1})e^{-x\left[
\theta (j+1)-t\right] }dx  \notag \\
&-B_{ij}\int\limits_{0}^{\infty }(x^{i}+x^{i+1})e^{-x\left[ \theta
(j+1)-t\right] }\left( 1-(1+\frac{\theta x}{\theta +1})e^{-\theta x}\right) \notag\\
&=\frac{A_{ig}\Gamma (i+1)}{\left( \theta (j+1)-t\right) ^{i+1}}\left[ 1+%
\frac{i+1}{\left( \theta (j+1)-t\right) }\right]  \notag \\
&-B_{ij}\left\{ \frac{\Gamma (i+1)}{\left( \theta (j+1)-t\right) ^{i+1}}%
\left[ 1+\frac{i+1}{\left( \theta (j+1)-t\right) }\right] \right.  \notag \\
&\left. -\frac{\Gamma (i+1)}{\left( \theta (j+2)-t\right) ^{i+1}}\left[ 1+%
\frac{i+1}{\left( \theta (j+2)-t\right) }\right] \right.  \notag \\
&\left. -\frac{\theta }{\theta +1}\left( \frac{\Gamma (i+2)}{\left( \theta
(j+2)-t\right) ^{i+2}}\right) \left[ 1+\frac{i+2}{\left( \theta
(j+2)-t\right) }\right] \right\}
\end{align}%
Which completes the proof.

\section{Distribution of the order statistics}

In this section, we derive closed form expressions for the pdfs of the $%
r_{th}$ order statistic of the $TLG$ distribution, also, the measures of
skewness and kurtosis of the distribution of the $r_{th}$ order statistic in
a sample of size $n$ for different choices of $n;r$ are presented in this
section. Let $X_{1},X_{2},...,X_{n}$ be a simple random sample from $(TLG)$
distribution with pdf and cdf given by \eqref{eq2.1} and \eqref{eq2.2}, respectively.

Let $X_{1},X_{2},...,X_{n}$ denote the order statistics obtained from this
sample. We now give the probability density function of $X_{r:n}$, say $%
f_{r:n}(x,\Phi )$ and the moments of $X_{r:n}$ $,r=1,2,...,n$. Therefore,
the measures of skewness and kurtosis of the distribution of the $X_{r:n}$
are presented. The probability density function of $X_{r:n}$ is given by%
\begin{equation}\label{eq4.1}
f_{r:n}(x,\Phi )=\frac{1}{B(r,n-r+1)}\left[ F(x,\Phi )\right] ^{r-1}\left[
1-F(x,\Phi )\right] ^{n-r}f(x,\Phi )
\end{equation}
where $F(x,\Phi )$ and $f(x,\Phi )$ are the cdf and pdf of the $(TLG)$
distribution given by \eqref{eq2.1}, \eqref{eq2.2}, respectively, and $B(.,.)$ is the
beta function, since $0<F(x,\Phi )<1$, for $x>0$, by using the binomial
series expansion of $\left[ 1-F(x,\Phi )\right] ^{n-r}$, given by
\begin{equation}\label{eq4.2}
\left[ 1-F(x,\Phi )\right] ^{n-r}=\sum\limits_{j=0}^{n-r}(-1)^{j}\binom{n-r%
}{j}\left[ F(x,\Phi )\right] ^{^{j}},
\end{equation}
we have
\begin{equation}\label{eq4.3}
f_{r:n}(x,\Phi )=\sum\limits_{j=0}^{n-r}(-1)^{j}\binom{n-r}{j}\left[
F(x,\Phi )\right] ^{r+j-1}f(x,\Phi ),
\end{equation}%
substituting from \eqref{eq2.1} and \eqref{eq2.2} into \eqref{eq4.3}, we can express the $k_{th}$
ordinary moment of the $r_{th}$ order statistics $X_{r:n}$ say $E(X_{r:n}^{k})$ as a liner combination of the $k_{th}$ moments of the $(TLG)$
distribution with different shape parameters. Therefore, the measures of
skewness and kurtosis of the distribution of $X_{r:n}$ can be calculated.

\section{Least Squares and Weighted Least Squares Estimators}

In this section we provide the regression based method estimators of the
unknown parameters of the transmuted Lindley geometric distribution, which
was originally suggested by Swain, Venkatraman and Wilson (1988) to estimate
the parameters of beta distributions. It can be used some other cases also.
Suppose $Y_{1},...,Y_{n}$ is a random sample of size $n$ from a distribution
function $G(.)$ and suppose $Y_{(i)}$; $i=1,2,...,n$ denotes the ordered
sample. The proposed method uses the distribution of $G(Y_{(i)})$. For a
sample of size $n$, we have%
\begin{align*}
E\left( G(Y_{(j)})\right) &=\frac{j}{n+1},V\left( G(Y_{(j)})\right) =\frac{%
j(n-j+1)}{(n+1)^{2}(n+2)} \\
\textrm{and }Cov\left( G(Y_{(j)}),G(Y_{(k)})\right) &=\frac{j(n-k+1)}{%
(n+1)^{2}(n+2)};\text{for\ }j<k\text{,}
\end{align*}
see Johnson, Kotz and Balakrishnan (1995). Using the expectations and the
variances, two variants of the least squares methods can be used.\medskip
\newline

\textbf{Method 1 (Least Squares Estimators)} . Obtain the estimators by
minimizing%
\begin{equation}\label{eq5.1}
\sum\limits_{j=1}^{n}\left( G(Y_{(j)}-\frac{j}{n+1}\right) ^{2},
\end{equation}%
with respect to the unknown parameters. Therefore in case of $TLG$
distribution the least squares estimators of $\theta ,p$ and $\lambda $ ,
say $,\widehat{\theta }_{LSE},$\ $\widehat{p}_{LSE}$ and $\widehat{\lambda }%
_{LSE}$ respectively, can be obtained by minimizing%
\begin{equation*}
\sum\limits_{j=1}^{n}\left[ \frac{1-(1+\frac{\theta x}{\theta +1}%
)e^{-\theta x}}{1-p(1+\frac{\theta x}{\theta +1})e^{-\theta x}}\left[
1+\lambda -\lambda \left( \frac{1-(1+\frac{\theta x}{\theta +1})e^{-\theta x}%
}{1-p(1+\frac{\theta x}{\theta +1})e^{-\theta x}}\right) \right] -\frac{j}{%
n+1}\right] ^{2}
\end{equation*}%
with respect to $\theta ,p$ and $\lambda $.\medskip \newline

\textbf{Method 2 (Weighted Least Squares Estimators).} The weighted least
squares estimators can be obtained by minimizing%
\begin{equation}\label{eq5.2}
\sum\limits_{j=1}^{n}w_{j}\left( G(Y_{(j)}-\frac{j}{n+1}\right) ^{2},
\end{equation}%
with respect to the unknown parameters, where%
\begin{equation*}
w_{j}=\frac{1}{V\left( G(Y_{(j)})\right) }=\frac{(n+1)^{2}(n+2)}{j(n-j+1)}.
\end{equation*}%
Therefore, in case of $TLG$ distribution the weighted least squares
estimators of $\theta ,p$ and $\lambda $ , say $,\widehat{\theta }_{WLSE},$\
$\widehat{p}_{WLSE}$and $\widehat{\lambda }_{WLSE}$ respectively\ , can be
obtained by minimizing%
\begin{equation*}
\sum\limits_{j=1}^{n}w_{j}\left[ \frac{1-(1+\frac{\theta x}{\theta +1}%
)e^{-\theta x}}{1-p(1+\frac{\theta x}{\theta +1})e^{-\theta x}}\left[
1+\lambda -\lambda \left( \frac{1-(1+\frac{\theta x}{\theta +1})e^{-\theta x}%
}{1-p(1+\frac{\theta x}{\theta +1})e^{-\theta x}}\right) \right] -\frac{j}{%
n+1}\right] ^{2}
\end{equation*}%
with respect to the unknown parameters only.

\section{Estimation and Inference}

In this section, we determine the maximum likelihood estimates (MLEs) of the
parameters of the $(TLG)$ distribution from complete samples only. Let $X_{1},X_{2},...,X_{n}$ be a random sample of size $n$  from $TLG$ $(\theta
,p,\lambda ,x)$.The likelihood function for the vector of parameters $\Phi=(\theta ,p,\lambda )$ can be written as%
\begin{align}\label{eq5.1}
Lf(x_{(i)},\Phi ) &=\Pi _{i=1}^{n}f(x_{(i)},\Phi )  \notag \\
&=\left( \frac{\theta ^{2}}{\theta +1}\right) ^{n}\text{ }(1-p)^{n}\Pi
_{i=1}^{n}(1+x_{i})\text{ }e^{-\theta \sum\limits_{i=1}^{n}x_{i}}\Pi
_{i=1}^{n}\left[ 1-p(1+\frac{\theta x_{i}}{\theta +1})e^{-\theta x_{i}}%
\right] ^{-2}  \notag \\
&\times \Pi _{i=1}^{n}\left\{ (1+\lambda )-2\lambda \left( \frac{1-(1+\frac{%
\theta x_{i}}{\theta +1})e^{-\theta x_{i}}}{1-p(1+\frac{\theta x_{i}}{\theta
+1})e^{-\theta x_{i}}}\right) \right\} .
\end{align}%
Taking the log-likelihood function for the vector of parameters $\Phi =(\theta ,p,\lambda )$ we get%
\begin{align}\label{eq5.2}
\ell=\log L &=2n\log \theta -n\log (1+\theta )+n\log
(1-p)+\sum\limits_{i=1}^{n}\log (1+x_{i})-\theta
\sum\limits_{i=1}^{n}x_{(i)}\notag\\
&-2\sum\limits_{i=1}^{n}\log \left[ 1-p(1+\frac{%
\theta x_{i}}{\theta +1})e^{-\theta x_{i}}\right]  \notag \\
&+\sum\limits_{i=1}^{n}\log \left\{ (1+\lambda )-2\lambda \left( \frac{%
1-(1+\frac{\theta x_{i}}{\theta +1})e^{-\theta x_{i}}}{1-p(1+\frac{\theta
x_{i}}{\theta +1})e^{-\theta x_{i}}}\right) \right\} .
\end{align}
The log-likelihood can be maximized either directly or by solving the
nonlinear likelihood equations obtained by differentiating \eqref{eq5.2}.
The
components of the score vector are given by
\begin{align}\label{eq5.3}
\frac{\partial \ell}{\partial p} &=\frac{-n}{1-p}%
+2\sum\limits_{i=1}^{n}\frac{(1+\frac{\theta x_{i}}{\theta +1})e^{-\theta
x_{i}}}{\left[ 1-p(1+\frac{\theta x_{i}}{\theta +1})e^{-\theta x_{i}}\right]
}  \notag \\
&-2\lambda \sum\limits_{i=1}^{n}\frac{\left[ 1-(1+\frac{\theta x_{i}}{
\theta +1})e^{-\theta x_{i}}\right] \left[ \frac{(1+\frac{\theta x_{i}}{%
\theta +1})e^{-\theta x_{i}}}{(1-p(1+\frac{\theta x_{i}}{\theta +1}%
)e^{-\theta x_{i}})^{2}}\right] }{\left\{ (1+\lambda )-2\lambda \left( \frac{%
1-(1+\frac{\theta x_{i}}{\theta +1})e^{-\theta x_{i}}}{1-p(1+\frac{\theta
x_{i}}{\theta +1})e^{-\theta x_{i}}}\right) \right\} }=0,
\end{align}%
`%
\begin{align}\label{eq5.4}
&\frac{\partial \ell}{\partial \theta } =\frac{2n}{\theta }-\frac{%
n}{1+\theta }-\sum\limits_{i=1}^{n}x_{i}-2p\sum\limits_{i=1}^{n}\frac{%
x_{i}e^{-\theta x_{i}}\left[ (1+\frac{\theta x_{i}}{\theta +1})-\frac{1}{%
\left( 1+\theta \right) ^{2}}\right] }{\left[ 1-p(1+\frac{\theta x_{i}}{%
\theta +1})e^{-\theta x_{i}}\right] }  \notag \\
&-2\lambda \sum\limits_{i=1}^{n}\frac{(1-p)}{\left\{ (1+\lambda )-2\lambda
\left( \frac{1-(1+\frac{\theta x_{i}}{\theta +1})e^{-\theta x_{i}}}{1-p(1+%
\frac{\theta x_{i}}{\theta +1})e^{-\theta x_{i}}}\right) \right\} }\left[
\frac{x_{i}e^{-\theta x_{i}}\left[ (1+\frac{\theta x_{i}}{\theta +1})-\frac{1%
}{\left( 1+\theta \right) ^{2}}\right] }{\left[ 1-p(1+\frac{\theta x_{i}}{%
\theta +1})e^{-\theta x_{i}}\right] ^{2}}\right] =0
\end{align}%
and%
\begin{equation}\label{eq5.5}
\frac{\partial \ell}{\partial \lambda }=\sum\limits_{i=1}^{n}%
\frac{1-2\left( \frac{1-(1+\frac{\theta x_{i}}{\theta +1})e^{-\theta x_{i}}}{%
1-p(1+\frac{\theta x_{i}}{\theta +1})e^{-\theta x_{i}}}\right) }{\left\{
(1+\lambda )-2\lambda \left( \frac{1-(1+\frac{\theta x_{i}}{\theta +1}%
)e^{-\theta x_{i}}}{1-p(1+\frac{\theta x_{i}}{\theta +1})e^{-\theta x_{i}}}%
\right) \right\} }=0.
\end{equation}%
We can find the estimates of the unknown parameters by maximum likelihood
method by setting these above non-linear equations \eqref{eq5.4}- \eqref{eq5.5} to zero and
solve them simultaneously. Therefore, we have to use mathematical package to get the MLE of the unknown
parameters.
Applying the usual large sample approximation, the MLE $\hat\Phi$ can be treated as being approximately trivariate normal  and variance-covariance matrix equal to the inverse of the expected information matrix, i.e.
$$
 \sqrt{n}(\hat\Phi - \Phi) \to N\left(0, nI^{-1}(\Phi)\right)\,,
$$
where $I^{-1}(\Phi)$ is the limiting variance-covariance matrix of $\hat\Phi$. The elements of the $3 \times 3$ matrix $I(\Phi)$ can be estimated by $I_{ij}(\hat\Phi) = -\ell_{\Phi_i\Phi_j} \vline_{\Phi = \hat{\Phi}}$, $i, j \in\{1, 2,3\}$.

Approximate two sided $100(1-\alpha)\%$ confidence intervals for $\theta, p$ and for $\lambda$ are, respectively, given by
$$
 \hat\theta \pm z_{\alpha/2}\sqrt{I^{-1}_{11}(\hat\theta)},
 \hat p \pm z_{\alpha/2}\sqrt{I^{-1}_{22}(\hat p)}$$
 and $$
    \hat\lambda \pm z_{\alpha/2}\sqrt{I^{-1}_{33}(\hat\lambda)},
    $$

where $z_{\alpha}$ is the upper $\alpha$th quantile of the standard normal distribution. Using \texttt{R} we can easily compute the Hessian matrix and its inverse and hence the standard errors and asymptotic confidence intervals.

\section{Application}

In this section, we use a real data set to show that the transmuted Lindley distribution can be a better model than one based on
the Lindley geometric  distribution and Lindley distribution. The data set given in Table 1 represents the waiting times (in minutes) before service of
100 bank customers.

\begin{table}[hbt]
 \caption{\label{Tab1} the waiting times (in minutes) before service of
100 bank customers.}
 \begin{small}
 \begin{center}
 \begin{tabular}{llllllllll}
 \hline
 0.8&0.8&1.3&1.5&1.8&1.9&1.9&2.1&2.6&2.7\\
2.9&3.1&3.2&3.3&3.5&3.6&4.0&4.1&4.2&4.2\\
4.3&4.3&4.4&4.4&4.6&4.7&4.7&4.8&4.9&4.9\\
5.0&5.3&5.5&5.7&5.7&6.1&6.2&6.2&6.2&6.3\\
6.7&6.9&7.1&7.1&7.1&7.1&7.4&7.6&7.7&8.0\\
8.2&8.6&8.6&8.6&8.8&8.8&8.9&8.9&9.5&9.6\\
9.7&9.8&10.7&10.9&11.0&11.0&11.1&11.2&11.2&11.5\\
11.9&12.4&12.5&12.9&13.0&13.1&13.3&13.6&13.7&13.9\\
14.1&15.4&15.4&17.3&17.3&18.1&18.2&18.4&18.9&19.0\\
19.9&20.6&21.3&21.4&21.9&23.0&27.0&31.6&33.1&38.5\\
 \hline
 \end{tabular}
 \end{center}
 \end{small}
\end{table}

\begin{table}[hbt]
 \caption{\label{Tab2} Estimated parameters of the Lindley,  Lindley geometric   and transmuted Lindley geometric  distribution for the waiting times (in minutes) before service of 100 bank customers.}
 \begin{small}
 \begin{center}
 \begin{tabular}{llll}
 \hline
 Model      & Parameter Estimate      & Standard Error & $-\ell(\cdot; x)$ \\
 \hline
 Lindley & $\hat{\theta} = 0.186$    & $0.013$       &  319.037\\\hline
Lindley  & $\hat{\theta} = 0.202$    & $0.034$       & 318.913\\
   Geometric & $\hat{p} = -0.242$ &  0.5270            &\\\hline
Transmuted & $\hat{\theta}= 0.171$ & 0.0351       & 317.207\\
Lindley  & $\hat{p} = 0.657 $       &  0.181      & \\
Geometric   &$\hat{\lambda}= -0.954$ &  0.192          &          \\
				
 \hline
 \end{tabular}
 \end{center}
 \end{small}
\end{table}

The variance covariance matrix of the MLEs under the transmuted  Lindley geometric  distribution is computed as
$$
 I(\hat\theta)^{-1}
 = \begin{pmatrix}
 0.001& -0.005& 0.002\\
-0.005 & 0.032& -0.020\\
0.002& -0.020&  0.037\\
\end{pmatrix}\,.
$$
Thus, the variances of the MLE of $\theta, p$ and $\lambda$ is $var(\hat\theta) = 0.0012,var(\hat p) = 0.0326,var(\hat a) =  0.0368.$ Therefore, $95\%$ confidence intervals for $\theta, p$ and $\lambda$ are $[0.102, 0.240],[ 0.302,1],$ and $[-0.577,  1]$ respectively.
\begin{table}[hbt]
 \caption{\label{Tab3} Criteria for comparison.}
 \begin{center}
 \begin{small}
 \begin{tabular}{llllllll}
 \hline
  Model               & K-S  & $-2\ell$ & AIC     & AICC    \\
  \hline
  Lindley           &0.0677  &  638.1 &640.1 &640.1   \\
 Lindley Geometric  & 0.0557   &   637.8 &641.8  &642  \\
  TLG             & 0.0017 & 634.414 &640.414  &640.664  \\
  \hline
 \end{tabular}
 \end{small}
 \end{center}
\end{table}

In order to compare the two distribution models, we consider criteria like K-S, $-2\ell$, AIC (Akaike information criterion)and AICC (corrected Akaike information criterion)  for the data set. The better distribution corresponds to smaller K-S, $-2\ell$, AIC and AICC  values:`
$$
 \mbox{AIC} = 2k - 2\ell\,, \quad \textrm{ and }\quad
 \mbox{AICC} = \mbox{AIC} + \frac{2k(k+1)}{n-k-1}\,,
$$

where $k$ is the number of parameters in the statistical model, $n$  the sample size and $\ell$ is the maximized value of the log-likelihood function under the considered model. Also, here for calculating the values of KS we use the sample estimates of $\theta, \alpha, a,b  $ and $c$. Table 2 shows the MLEs under both distributions, Table 3 shows the values of K-S, $-2\ell$, AIC and AICC values. The values in table 3 indicate that the transmuted Lindley geometric  distribution leads to a better fit than the  Lindley geometric  distribution and Lindely distribution.

A density plot compares the fitted
densities of the models with the empirical histogram of the observed data (Fig. 4).
The fitted density for the transmuted Linldey geometric  model is closer to the empirical histogram than the
fits of the Lindley geometric  and Lindley sub-models.

\begin{figure}[H]
\centering
  \includegraphics[width=8.0cm]{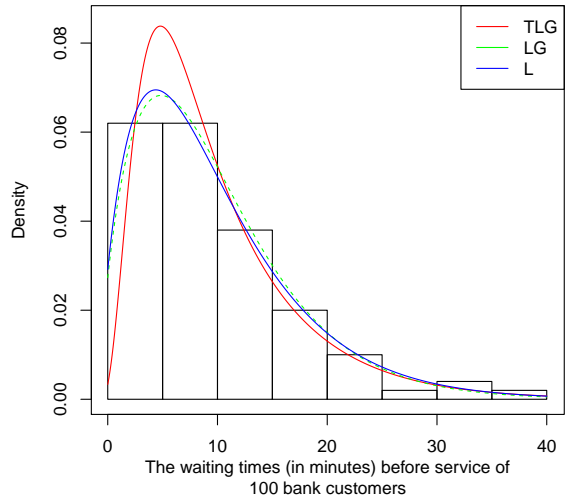}
 \caption{Estimated densities of the models for  the waiting times (in minutes) before service of
100 bank customers.\label{fig4.pdf}}
 \end{figure}

\begin{figure}[H]
 \begin{center}
\includegraphics[width=8cm]{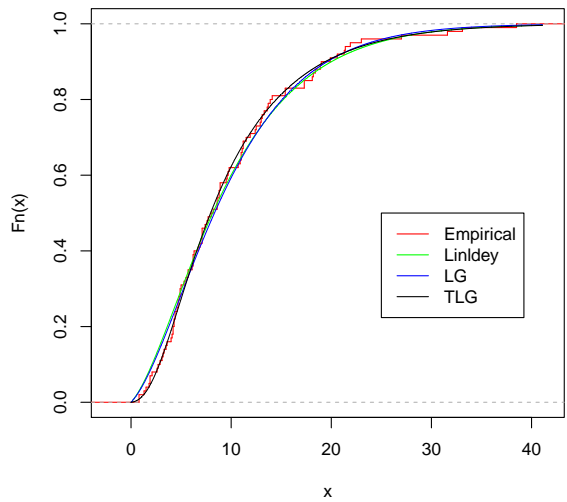}
\caption{Empirical, fitted Lindley, Lindley geometric  and transmuted Lindley geometric cdf of the the waiting times (in minutes) before service of
100 bank customers. \label{figaaaaaq.pdf}}
\end{center}
\end{figure}

\section{Conclusion}

Here we propose a new model, the so-called the transmuted  Lindley geometric  distribution which extends the
 Lindley geometric distribution in the analysis of data with real support. An obvious reason for generalizing a standard distribution is because the generalized form provides larger flexibility in modeling real data. We derive expansions for  moments and for the moment generating function. The estimation of parameters is approached by the method of maximum likelihood, also the information matrix is derived.  An application of the transmuted Lindley geometric distribution to real data show that the new distribution can be used quite effectively to provide better fits than Lindley geometric and  Lindley distribution.
\medskip \newline


\begin{thebibliography}{99}

\bibitem{2} Adamidis K.,  Dimitrakopoulou T., and Loukas S. , On a generalization of the
exponential-geometric distribution, Statist. Probab. Lett. 73 (2005), pp. 259-269.
\bibitem{adamis} Adamidis, K., and Loukas, S. (1998). A lifetime distribution with decreasing failure rate. Statistics and Probability Letters, 39(1), 35-42.
\bibitem{bakouch} Bakouch, H. S., Al-Zahrani, B. M., Al-Shomrani, A. A., Marchi, V. A., and Louzada, F. (2012). An extended Lindley distribution. Journal of the Korean Statistical Society, 41(1), 75-85.


        \bibitem{barreto5} Barreto-Souza, W., and Cribari-Neto, F. (2009). A generalization of the exponential-Poisson distribution. Statistics and Probability Letters, 79(24), 2493-2500.
        \bibitem{6} Barreto-Souza, W., de Morais, A. L., and Cordeiro, G. M. (2011). The Weibull-geometric distribution. Journal of Statistical Computation and Simulation, 81(5), 645-657.
         \bibitem{7} Cancho, V. G., Louzada-Neto, F., and Barriga, G. D. (2011). The Poisson-exponential lifetime distribution. Computational Statistics and Data Analysis, 55(1), 677-686.
           \bibitem{chahkandi} Chahkandi, M., and Ganjali, M. (2009). On some lifetime distributions with decreasing failure rate. Computational Statistics and Data Analysis, 53(12), 4433-4440.
    \bibitem{gitany} Ghitany, M. E., Atieh, B., and Nadarajah, S. (2008). Lindley distribution and its application. Mathematics and Computers in Simulation, 78(4), 493-506.
        \bibitem{kus} Kus, C. (2007). A new lifetime distribution. Computational Statistics and Data Analysis, 51(9), 4497-4509.
    \bibitem{mahmuti} Mahmoudi, E., and Zakerzadeh, H. (2010). Generalized Poisson–Lindley distribution. Communications in Statistics—Theory and Methods, 39(10), 1785-1798.
          \bibitem{18} Mahmoudi, E., and Torki, M. (2011). Generalized inverse Weibull-Poisson distribution and its applications. Submited to Journal of Statistical Computation and Simulation.
               \bibitem{17} Mahmoudi, E., and Sepahdar, A. (2011). Exponentiated Weibull-Poisson distribution and its applications. Submited to Mathematics and Computer in Simulation.
            \bibitem{16} Mahmoudi, E., and Jafari, A. A. (2012). Generalized exponential–power series distributions. Computational Statistics and Data Analysis, 56(12), 4047-4066.
         \bibitem{marshall} Marshall, A. W., and Olkin, I. (1997). A new method for adding a parameter to a family of distributions with application to the exponential and Weibull families. Biometrika, 84(3), 641-652.

         \bibitem{faton1}Merovci, F.,(2013). Transmuted Rayleigh distribution. Austrian Journal of Statistics, Volume 42, Number 1, 21–31.
                        \bibitem{faton2} Merovci, F.,(2013). Transmuted generalized Rayleigh distribution. Journal of Statistics Applications and Probability, Volume 2,No. 3, 1-12.
                        \bibitem{faton3} Merovci, F.,(2013). Transmuted Lindley distribution. International Journal of Open
Problems in Computer Science and Mathematics, Volume 6, No. 2, 63-72.

\bibitem{20} Morais, A. L., and Barreto-Souza, W. (2011). A compound class of Weibull and power series distributions. Computational Statistics and Data Analysis, 55(3), 1410-1425.

\bibitem{lindley} Lindley, D. V. (1958). Fiducial distributions and Bayes' theorem. Journal of the Royal Statistical Society. Series B (Methodological), 102-107.
    \bibitem{lindley2} Lindley, D. V. (1965). Introduction to probability and statistics from bayesian viewpoint. part 2 inference. CUP Archive.
   \bibitem{14} Louzada, F., Roman, M., and Cancho, V. G. (2011). The complementary exponential geometric distribution: Model, properties, and a comparison with its counterpart. Computational Statistics and Data Analysis, 55(8), 2516-2524.



   \bibitem{sankaran} Sankaran, M. (1970). The Discrete Poisson-Lindley Distribution. Biometrics, 145-149.

    \bibitem{tahmasbi} Tahmasbi, R., and Rezaei, S. (2008). A two-parameter lifetime distribution with decreasing failure rate. Computational Statistics and Data Analysis, 52(8), 3889-3901.




\bibitem{2012} Zakerzadeh, H. and Mahmoudi, E. (2012). A new two parameter lifetime distribution: model
and properties. arXiv:1204.4248 [stat.CO].

     \bibitem{zaker} Zakerzadeh, H., and Dolati, A. (2009). Generalized lindley distribution. Journal of Mathematical Extension, 3(2), 13-25.


\end{thebibliography}
\end{document}